\documentclass[conference]{IEEEtran}
\IEEEoverridecommandlockouts

\usepackage{graphicx}
\usepackage{textcomp}
\usepackage{xcolor}
\usepackage{flushend}
\usepackage[caption=false,font=footnotesize]{subfig}
\usepackage{multirow}
\usepackage{array}

\usepackage{amsmath}
\usepackage{amssymb}
\usepackage{amsfonts}

\usepackage{amsthm}

\usepackage{algorithm}
\usepackage{algpseudocode}   

\usepackage{graphicx}

\usepackage{bm}              
\usepackage{mathtools}       

\usepackage{cite}            

\theoremstyle{plain}

\theoremstyle{remark}

\def\BibTeX{{\rm B\kern-.05em{\sc i\kern-.025em b}\kern-.08em
    T\kern-.1667em\lower.7ex\hbox{E}\kern-.125emX}}
    
\begin{document}

\title{LarS-Net: A Large-Scale Framework for Network-Level Spectrum Sensing}

\author{
\IEEEauthorblockN{Hao Guo\IEEEauthorrefmark{1}, 
Ruoyu Sun\IEEEauthorrefmark{1},
Amir Hossein Fahim Raouf\IEEEauthorrefmark{2},
Rahil Gandotra\IEEEauthorrefmark{1},
Jiayu Mao\IEEEauthorrefmark{3},
Mark Poletti\IEEEauthorrefmark{1}}

\IEEEauthorblockA{\IEEEauthorrefmark{1}Department of Wireless Technologies, CableLabs, Louisville, Colorado, USA}
\IEEEauthorblockA{\IEEEauthorrefmark{2}Department of Electrical and Computer Engineering, North Carolina State University, Raleigh, NC, USA}
\IEEEauthorblockA{\IEEEauthorrefmark{3}Department of Electrical and Computer Engineering, The Ohio State University, Columbus, Ohio, USA}
Email: \{h.guo, r.sun, r.gandotra, m.poletti\}@cablelabs.com, amirh.fraouf@ieee.org, mao.518@osu.edu

\thanks{This paper is under review at an IEEE conference.}
}

\maketitle

\begin{abstract} 
As the demand of wireless communication continues to rise, the radio spectrum—a finite resource—requires increasingly efficient utilization. This trend is driving the evolution from static, stand-alone spectrum allocation toward spectrum sharing and dynamic spectrum sharing. A critical element of this transition is spectrum sensing, which facilitates informed decision-making in shared environments. Previous studies on spectrum sensing and cognitive radio have been largely limited to individual sensors or small sensor groups. In this work, a large-scale spectrum sensing network (LarS-Net) is designed in a cost-effective manner. In this framework, spectrum sensors are either co-located with cellular network base stations (BSs) to share the tower, backhaul, and power infrastructure, or integrated directly into BSs as a new feature leveraging active BS antenna systems. As an example incumbent system, fixed service microwave link operating in the lower-7 GHz band and deployed in the same geographic area as LarS-Net is investigated. This band is a primary candidate for 6G, currently under consideration by the WRC-23, ITU, and FCC, where sharing with incumbent systems is anticipated. Based on Monte Carlo simulations, we determine the minimum subset of BSs equipped with sensing capability to guarantee a target incumbent detection probability. The simulations account for various sensor antenna configurations, propagation channel models, and duty cycles for both incumbent transmissions and sensing operations. Building on this framework, we introduce three network-level sensing performance metrics: Emission Detection Probability (EDP), Temporal Detection Probability (TDP), and Temporal Mis-detection Probability (TMP), which jointly capture spatial coverage, temporal detectability, and multi-node diversity effects. Using these metrics, we analyze the impact of LarS-Net inter-site distance, noise uncertainty, and sensing duty-cycle on large-scale sensing performance. The results demonstrate that sensing capabilities on only 4\% to 25\% of BSs are sufficient to achieve a 90\% EDP in an urban macrocell environment, highlighting the feasibility of infrastructure-integrated large-scale sensing.
\end{abstract}

\section{Introduction}
The rapid growth of mid-band and high-band wireless deployments has intensified the need for reliable spectrum-awareness mechanisms to ensure safe coexistence between heterogeneous radio systems. While spectrum sensing has been extensively studied in the context of cognitive radio, most classical techniques operate at an individual receiver and focus primarily on physical-layer detection of incumbent signals. Such link-centric designs provide limited insight into how a distributed network of base stations~(BSs) jointly senses incumbent activity, implements protective mechanisms, and adapts its operation across wide geographic areas. This limitation is particularly relevant for fixed service~(FS) links operating in the lower 7~GHz band, which are widely deployed for long-range, high-reliability communications. Moreover, this gap is becoming increasingly critical as 6G systems are expected to provide wide-area spectrum sensing and scanning capabilities, as newly adopted in 3GPP TR~22.870~\cite{3gpp_tr_22_870} in Nov.~2025.

Prior work on cooperative and distributed spectrum sensing has primarily aimed to enhance the robustness of single-node detection. Early contributions investigated sparse distributed recovery~\cite{Bazerque2010}, compressive wideband sensing~\cite{Zeng2011}, and diffusion-based fusion strategies~\cite{Trigka2022}. More recently, cross-layer and protocol-level sensing frameworks have emerged, including cross-layer evaluation systems for cognitive-radio sensing and allocation~\cite{zhang2024cross} as well as energy-detection–based sensing designs developed for mmWave and THz communications~\cite{zang2023spectrum}. Machine learning~(ML)-based cooperative sensing has also been investigated, for example through hybrid deep models for congestion-aware spectrum prediction and resource allocation in cognitive-radio internet-of-vehicle networks~\cite{ramsha2022_HybridML_IoV}. These approaches enhance detection sensitivity but largely focus on algorithmic fusion or data-driven reconstruction. They rarely incorporate deployment-specific factors such as BS density (e.g., inter-site distance~(ISD)), 3D antenna patterns, realistic sectorization, or propagation-driven spatial effects. Consequently, existing methods cannot characterize network-level sensing behavior, including wide-area detection coverage, density-dependent sensitivity, missed-airtime patterns, or scaling trends as BS topology varies. 

In parallel, integrated sensing and communication~(ISAC) has emerged as an active research direction for 6G. The majority of ISAC studies focus on joint radar–communication signal design, such as beamforming or covariance optimization, exemplified by~\cite{Xu2023_ISAC_DAN}, dual-functional radar–communication waveform design~\cite{Liu2022_ISAC_Mag}, and ISAC under distributed multiple-input multiple-output~(D-MIMO)~\cite{guo2025dmimo}. These works aim at high-resolution target detection, localization, or radar imaging rather than wide-area spectrum scanning or incumbent detection. Meanwhile, measurement-driven studies such as~\cite{raouf2025} and coexistence analyses such as~\cite{sun2024icc} have provided valuable insights into long-term spectrum occupancy and mid-band interference, but they do not quantify distributed sensing performance nor its dependence on cellular geometry.

Overall, although cooperative sensing, ML-assisted spectrum prediction, and ISAC have each advanced significantly, no existing approach offers a deployment-aware framework capable of quantifying cell topology-dependent sensing coverage in realistic sub-6~GHz networks. Existing literature lacks models that jointly incorporate 3D antenna patterns, ISD, street-level propagation, load-dependent interference, and multi-cell spatial geometry. This gap motivates our study, which introduces a deployment-aware distributed sensing evaluation framework that directly relates network geometry to wide-area sensing performance and detection coverage.

This paper addresses critical research gaps in existing literature by developing a comprehensive modeling framework for distributed large-scale spectrum sensing networks~(LarS-Net). The proposed framework evaluates the detectability of emitters by assuming that either all BSs within a cellular network, or a strategic subset thereof, are equipped with spectrum sensing capabilities. While the LarS-Net architecture is designed to be agnostic to the incumbent system type, this work utilizes an FS link at 7.25 GHz as the primary use case to validate the framework's efficacy in a high-priority 6G candidate band. Our main contributions are as follows.

\begin{itemize}
    \item \textbf{Deployment-aware spectrum-sensing simulator:}
    We develop a scalable simulator that integrates cellular geometry, multi-sector BS antennas, realistic directional incumbent patterns, free-space and Longley-Rice propagation model, time-varying incumbent activity, measurement noise, and BS sensing duty cycles. This simulator captures spatial sensing behavior on city-wide networks. 

    \item \textbf{Novel network-level sensing metrics:}
    We introduce three metrics: Emission Detection Probability~(EDP), Temporal Detection Probability~(TDP), and Temporal Mis-detection Probability~(TMP) that jointly quantify spatial coverage, temporal detection consistency, and network-level failure modes.

    \item \textbf{ISD analysis and comparison between LarS-Net and single sensor:}
    Using the proposed metrics, we analyze how sensing performance scales with LarS-Net ISD and equivalently BS density, revealing saturation and degradation trends. Moreover, we compare our sensing network with a single sensor in different incumbent dropping areas to demonstrate the effectiveness of the LarS-Net.
\end{itemize}

To the best of our knowledge, this is the first work to characterize network-level spectrum sensing performance under realistic large-scale deployments and to integrate these sensing metrics into an ISAC-style optimization framework.

\section{System Model and LarS-Net Simulator}
\label{sec:system}
We consider a large-scale cellular deployment where a network of BSs (or sensors) senses the signal from an incumbent transmitter. The proposed LarS-Net simulator models the spatial geometry, antenna patterns, and the propagation environment, and computes the received power at each sensor under different channel models. 

\begin{figure}[t]
    \centering
    \includegraphics[width=1\columnwidth]{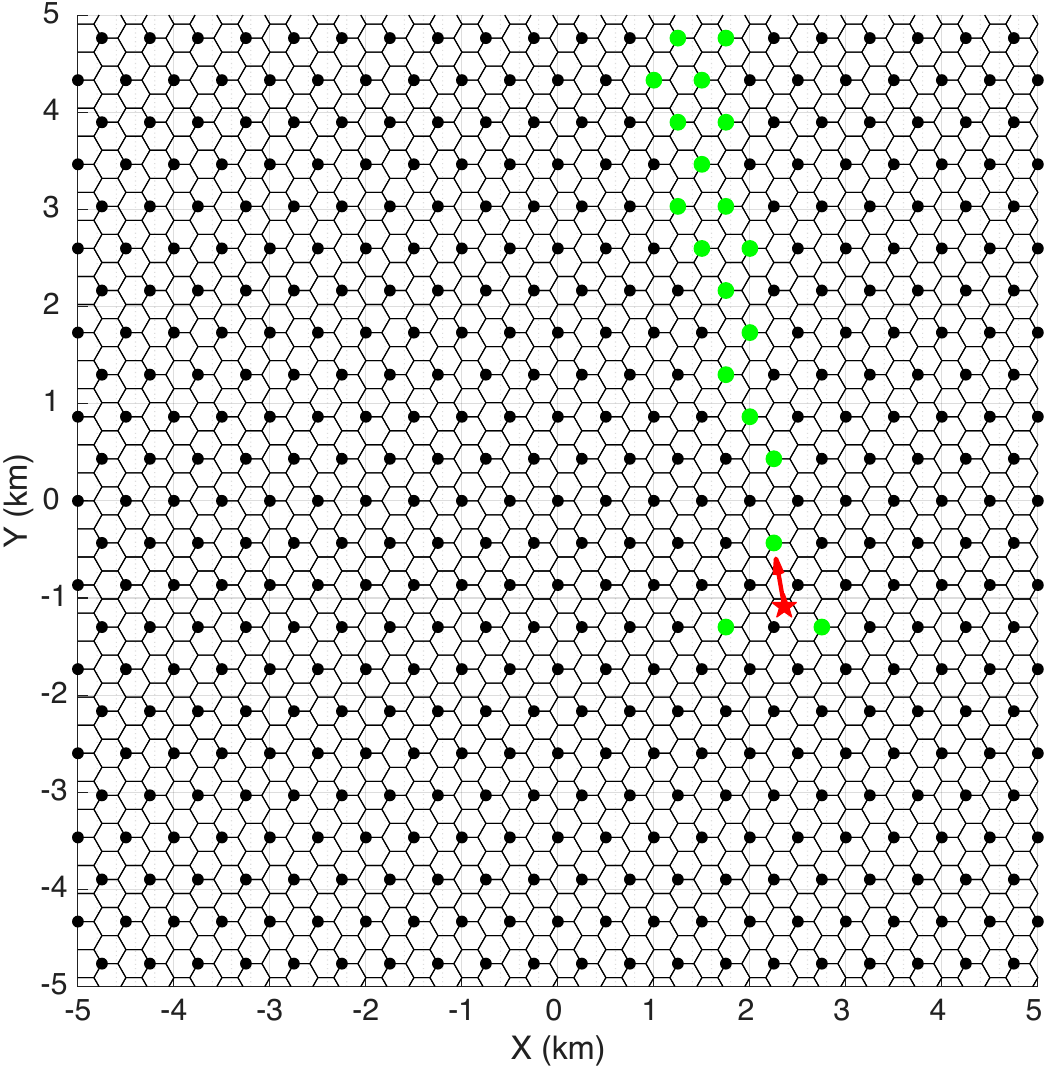}
    \caption{System model with $D_{\text{ISD}}$  = 500~m and free-space propagation. Black dots are sensors with 3-sector antenna and the red star represents the incumbent with an arrow pointing at its transmit direction. Sensors exceeding the detection threshold are shown in green.}
    \label{fig:model}
\end{figure}

\subsection{Network Geometry and Antenna Configuration}
The simulation area is configurable, and we consider a 10\,\text{km}$ \times 10\,\text{km}$ square centered at the origin, with coordinates $(x,y) \in [-L/2,L/2]^2$. As illustrated in Fig.~\ref{fig:model}, a regular hexagonal cellular layout with tri-sector BSs is deployed with ISD denoted as $D_{\text{ISD}}$. $D_{\text{ISD}} = 500\,\text{m}$ in the 7 GHz band in macro cellular networks \cite{3gpp_tr_38_914} \cite{3gpp_pcr_38_914}. Let $\mathcal{B}$ denote the set of BS sites. Each site is located at $(x_i,y_i,h_{\text{BS}})$ with BS height $h_{\text{BS}} $ and carries three sectors
with azimuth boresight orientations $\boldsymbol{\phi}_{\text{sec}} = [60^\circ, 180^\circ, 300^\circ]$. The azimuth angle $\phi$ is defined in a counter-clockwise direction with $\phi = 0^\circ$ pointing to the positive $x$-axis and $\phi = 90^\circ$ pointing to the positive $y$-axis. 

To minimize deployment cost, LarS-Net leverages existing cellular network infrastructure without constructing new sites. Spectrum sensors are co-located with all BSs or a selected subset of BSs. Our simulation evaluates the minimum number of sensors, equivalently, the maximum allowable LarS-Net ISD, required to achieve a certain incumbent detection probability.

\begin{figure*}[t]
\centering

\subfloat[FSPL]{%
  \begin{minipage}[t]{0.32\textwidth}
    \centering
    \includegraphics[width=\linewidth]{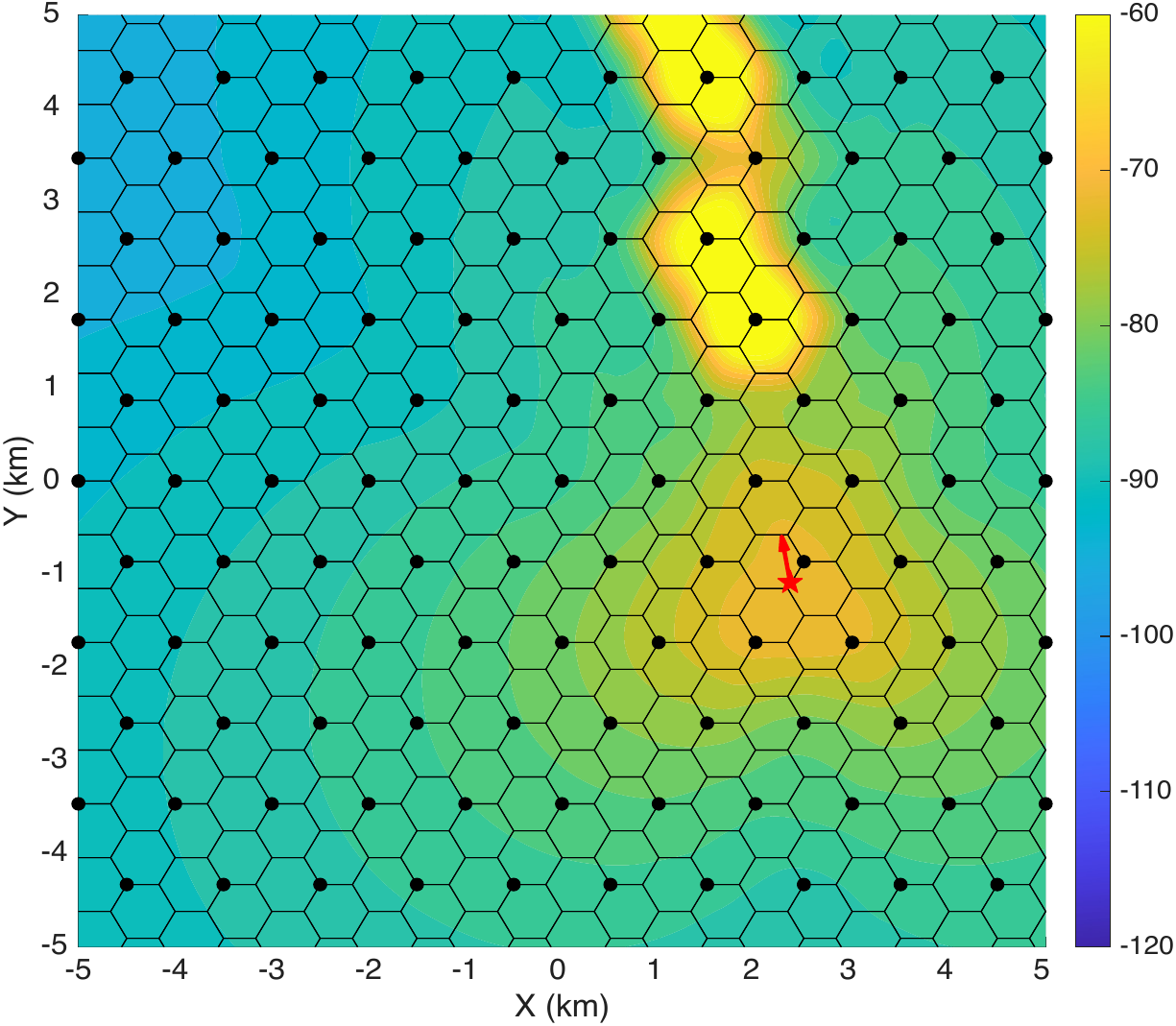}
  \end{minipage}
}\hfill
\subfloat[LRM (Boulder, CO)]{%
  \begin{minipage}[t]{0.32\textwidth}
    \centering
    \includegraphics[width=\linewidth]{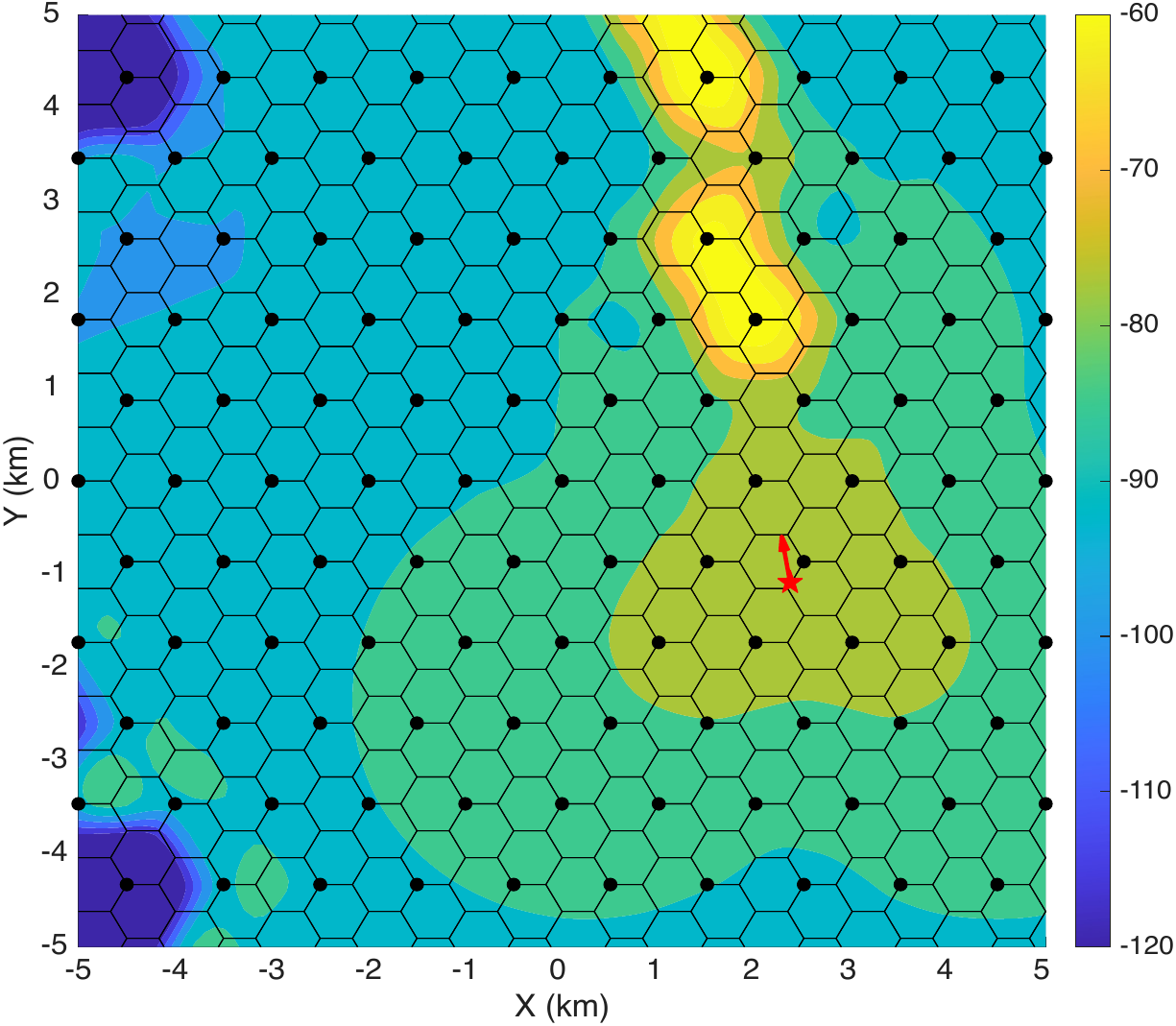}
  \end{minipage}
}\hfill
\subfloat[Google Earth (Boulder, CO)]{%
  \begin{minipage}[t]{0.32\textwidth}
    \centering
    \includegraphics[width=0.92\linewidth]
    {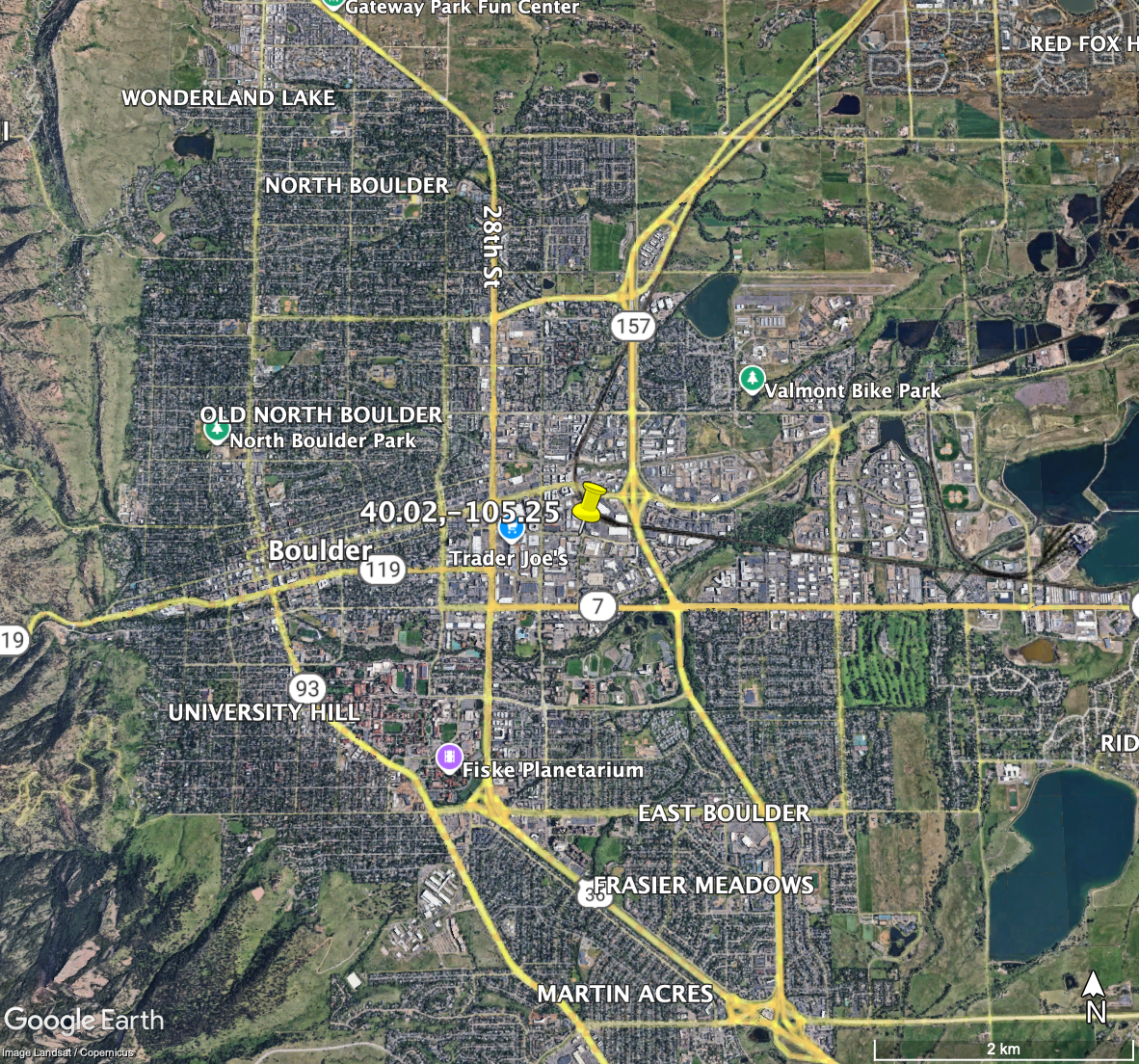}
  \end{minipage}
}

\caption{Spatial received-power maps under (a) FSPL, (b) LRM at Boulder, CO, and (c) terrain map of Boulder, CO from Google Earth.  Black markers show BS positions and red stars denote the incumbent transmitter.}
\label{fig:heatmaps_models}
\end{figure*}

An incumbent FS Microwave transmitter is placed at a random position $(x_{\text{In}}, y_{\text{In}}, h_{\text{In}})$ within the simulated area, following a uniform distribution.  The azimuth orientation of the directional FS incumbent antenna is also randomized with a uniform distribution. 

Two types of sensing antenna configurations are considered. The first configuration employs a sectored antenna based on the BS's active antenna system (AAS). This antenna forms a wide azimuth beam covering a 120$^\circ$ sector, combined with a narrow vertical beam that provides antenna gain and improved sensitivity. The antenna radiation pattern follows the 3GPP TR 38.921 model \cite{3gpp_tr_38_921}. This sectored sensing configuration requires tight integration between LarS-Net and the cellular communication network. The second sensing configuration employs an omnidirectional antenna that follows the ITU-R F.1336 model\cite{ITU1336}. In this case, a stand-alone sensor is co-located with the BS, sharing the same tower, backhaul, and power infrastructure, but operating independently of the cellular antenna system.



\subsection{Propagation Channel Models}

For each emitter-sensor pair, the simulator computes the path loss $L_i$ (dB). Since both the FS transmitter and the BS-hosted sensors are positioned above the clutter height (e.g., buildings and vegetation), the framework primarily employs free space path loss (FSPL) and the Longley-Rice model (LRM). To incorporate terrain-dependent effects, the LRM is implemented using MATLAB's \texttt{propagationModel('longley-rice')} interface. The simulation region is anchored at a reference geodetic coordinate $(\text{lat}_0,\text{lon}_0)$ corresponding to the southwest corner. Local $(x,y)$ coordinates are converted to latitude and longitude using the East-North-Up (ENU) to geodetic transformation, and then \texttt{txsite}/\texttt{rxsite} objects are created for the incumbent and each BS. The LRM accounts for ground conductivity, permittivity, atmospheric refractivity, and climate zone parameters, producing distance- and terrain-dependent path loss $L_i^{\text{LRM}}$.




\subsection{Received Power and Detection Threshold}
Given a selected path-loss model $L_i$, the received power in dBm at the $i^{\text{th}}$ BS is expressed as:
\begin{equation}
    P_{\text{rx},i} = P_{\text{EIRP,max}} - G_{\text{I,max}} + G_{\text{I}}(\phi_i,\theta_i) - L_i + G_{\text{BS},i}\,
\end{equation}
where $P_{\text{EIRP,max}}$ denotes the emitter's max Effective Isotropic Radiated Power (EIRP) in dBm, $G_{\text{I,max}}$ is the emitter's max antenna gain in dBi, $G_{\text{I}}(\phi_i,\theta_i)$ is the emitter's antenna gain toward the $i^{\text{th}}$ BS,  and $G_{\text{BS},i}$ represents the antenna gain of the $i^{\text{th}}$ BS in the direction of the emitter. To compare this against a power spectral density (PSD) threshold in dBm/MHz, the received power density is derived as
\begin{equation}\label{eq:conv_MHz}
    PSD_{\text{rx},i} 
    = P_{\text{rx},i} + 10\log_{10}\!\left(\frac{B}{1\,\text{MHz}}\right),
\end{equation}
where $B$ is the emitter's transmission bandwidth. A BS is considered to be \emph{above threshold} if its received power density satisfies ${PSD_{\text{rx},i} \ge \gamma_{\text{th}}}$, where $\gamma_{\text{th}}$ denotes the prescribed energy-detection threshold in dBm/MHz.

\subsection{Visualization of LarS-Net Received-Power Maps}
For each propagation model (FSPL and LRM), the LarS-Net simulator generates two types of visualizations:
\begin{enumerate}
    \item An interpolated heatmap of received power over the
          entire $L \times L$ area, highlighting the main sensing
          lobe and spatial decay of the incumbent signal.
    \item A binary threshold map where BSs above the sensing
          threshold are marked in a different color, directly
          visualizing which parts of the network can reliably
          sense the incumbent (e.g., Fig.~\ref{fig:model}).
\end{enumerate}
For example, Fig.~\ref{fig:heatmaps_models} (a) and Fig.~\ref{fig:heatmaps_models}(b) illustrate the interpolated received power heatmaps under the FSPL and LRM propagation models, respectively. Fig.~\ref{fig:heatmaps_models}(c) demonstrates the terrain profile of the investigated region in Boulder, Colorado, USA, where mountainous terrain is visible on the western (left) side of the simulation area. As observed in Fig.~\ref{fig:heatmaps_models}(b) , the LRM introduces terrain- and environment-induced shadowing and diffraction effects. These factors result in a non-monotonic sensing footprint across the BS locations, contrasting with the idealized radial symmetry of the FSPL model.

\section{Proposed Network-Level Sensing Metrics}\label{sec:metrics}

In this section, we formalize the time-slot-based sensing model used in LarS-Net and introduce three network-level sensing
metrics: EDP, TDP, and TMP.

\subsection{Time-Slot and Detection Model}
Time is partitioned into sensing slots indexed by ${t = 1,\dots,T}$, each having a fixed duration. The slot length is chosen to exceed the integration time required for reliable energy detection, ensuring that each slot supports a complete sensing decision.

The incumbent's transmission state in slot $t$ is modeled as a Bernoulli random variable $a_t \in \{0,1\}$ with
\begin{align}
  \Pr(a_t = 1) = p_{\text{on}},  
\end{align}
where $a_t = 1$ denotes that the incumbent is active during slot $t$.
The set of active slots is defined as 
\begin{align}
    \mathcal{T}_{\text{on}} = \{\, t : a_t = 1 \,\},
\end{align}
and its cardinality is given by $N_{\text{on}} = |\mathcal{T}_{\text{on}}|$.

For each emitter-sensor pair and slot $t$, the LarS-Net simulator computes the received power over the full system bandwidth $B$, denoted by $P_{r,i}[t]$ (in dBm), based on the geometry and propagation models described in Section~\ref{sec:system}. This
quantity is converted into a per-megahertz value using \eqref{eq:conv_MHz}.

Given a sensing threshold $\gamma_{\text{th}}$ (in dBm/MHz), the binary per-BS detection indicator under ideal listening (i.e., all BSs operate in sensing mode) is defined by
\begin{equation}
    d_{i,t}^{\text{I}} =
    \mathbf{1}\big\{PSD_{\text{rx},i}[t] \ge \gamma_{\text{th}}\big\},
\end{equation}
with ``I" refers to ``ideal". In this paper, we assume independent sensing activity across BSs and slots; correlated duty-cycling is left for future work. Moreover, a network-level detection is declared whenever at least $K$ sensors exceed the threshold
\begin{equation}
    D_t^{\text{I}} =
    \mathbf{1}\!\left\{\sum_{i\in\mathcal{B}} d_{i,t}^{\text{I}} \ge K\right\}.
\end{equation}

In practice, each BS senses only a fraction of the time due to communication requirements. We model the duty cycle by  $d_s \in [0,1]$ and introduce an independent Bernoulli activity indicator $b_{i,t}\sim\mathrm{Bernoulli}(d_s)$
for each BS and slot, where $b_{i,t}=1$ if BS $i$ is in sensing mode in slot $t$. The effective detection indicator then becomes $d_{i,t} = d_{i,t}^{\text{I}}\, b_{i,t}$, and the actual network decision considering duty cycle is defined as
\begin{equation}
    D_t =
    \mathbf{1}\!\left\{\sum_{i\in\mathcal{B}} d_{i,t} \ge K\right\}.
\end{equation}

\subsection{Emission Detection Probability}

The \textit{Emission Detection Probability} (EDP) quantifies the
intrinsic sensing capability of a deployment when all BSs
continuously operate in sensing mode. It is defined as the
probability that a sensor or sensing network correctly detects the incumbent
during an ON slot under ideal listening (duty cycle is 100\%)
\begin{equation}
    \text{EDP}
    \triangleq
    \Pr\big(D_t^{\text{I}} = 1 \,\big|\, a_t = 1\big).
        \label{eq:EDPdef}
\end{equation}
In the simulator, EDP is estimated empirically as
$\widehat{\text{EDP}}=\frac{1}{N_{\text{on}}}
    \sum_{t\in\mathcal{T}_{\text{on}}} D_t^{\text{I}}$, which averages the ideal-listening detection outcomes over all
incumbent-active slots. EDP reflects the joint influence of BS
density, antenna patterns, and propagation characteristics, and
serves as an upper bound on any network-level sensing
performance achievable under duty-cycled or time-shared
operation.

\subsection{Temporal Detection Probability (TDP)}
The \emph{Temporal Detection Probability} (TDP) quantifies the fraction of
incumbent airtime that is effectively protected by the network
under a sensing duty cycle $d_s$. It is defined as
\begin{equation}
    \text{TDP}(d_s)
    \triangleq
    \Pr\big(D_t = 1 \,\big|\, a_t = 1\big),
    \label{eq:TDPdef}
\end{equation}
and estimated in LarS-Net by $\widehat{\text{TDP}}(d_s)=\frac{1}{N_{\text{on}}}\sum_{t\in\mathcal{T}_{\text{on}}} D_t$, which averages the network-level detection decisions over all
incumbent-active slots. Comparing (\ref{eq:EDPdef}) and (\ref{eq:TDPdef}), TDP depends on the duty cycle and satisfies
\begin{equation}
    0 \le \text{TDP}(d_s) \le \text{EDP}, \quad
    \forall d_s \in [0,1],
\end{equation}
and TDP = EDP when duty cycle = 1. TDP therefore generalizes the classical notion of detection
probability to a network-level, time-aware metric of effective incumbent detection.

\subsection{Temporal Mis-detection Probability (TMP)}
The \emph{Temporal Mis-detection Probability} (TMP) characterizes the
fraction of time during which the incumbent is active but not
successfully detected by any sensor. We distinguish between a conditional form, evaluated only over incumbent-active slots, and an absolute form, evaluated over all time slots.

The conditional TMP with incumbent on is defined as
\begin{equation}
    \text{TMP}_{\text{ON}}(d_s)
    \triangleq
    \Pr\big(D_t = 0 \,\big|\, a_t = 1\big)
    = 1 - \text{TDP}(d_s),
    \label{eq:hap_on_def}
\end{equation}
while the absolute TMP over all slots is
\begin{equation}
    \text{TMP}_{\text{abs}}(d_s)
    \triangleq
    \Pr(a_t = 1,\, D_t = 0) = p_{\text{on}}\, \text{TMP}_{\text{ON}}(d_s).
\end{equation}

In the simulator, the absolute TMP is estimated through
$\widehat{\text{TMP}}_{\text{abs}}(d_s) = \frac{1}{T}\sum_{t=1}^{T} a_t \big(1 - D_t\big)$.

In particular, values near one indicate fully reliable sensing (EDP or TDP) or complete mis-detection (TMP), while values near zero correspond respectively to a non-detectable deployment (EDP), no effective detection under duty-cycling (TDP), or negligible mis-detection (TMP). 





\begin{figure*}[t]
\centering

\subfloat[ISD = 500 m]{%
  \begin{minipage}[t]{0.32\textwidth}
    \centering
    \includegraphics[width=\linewidth]{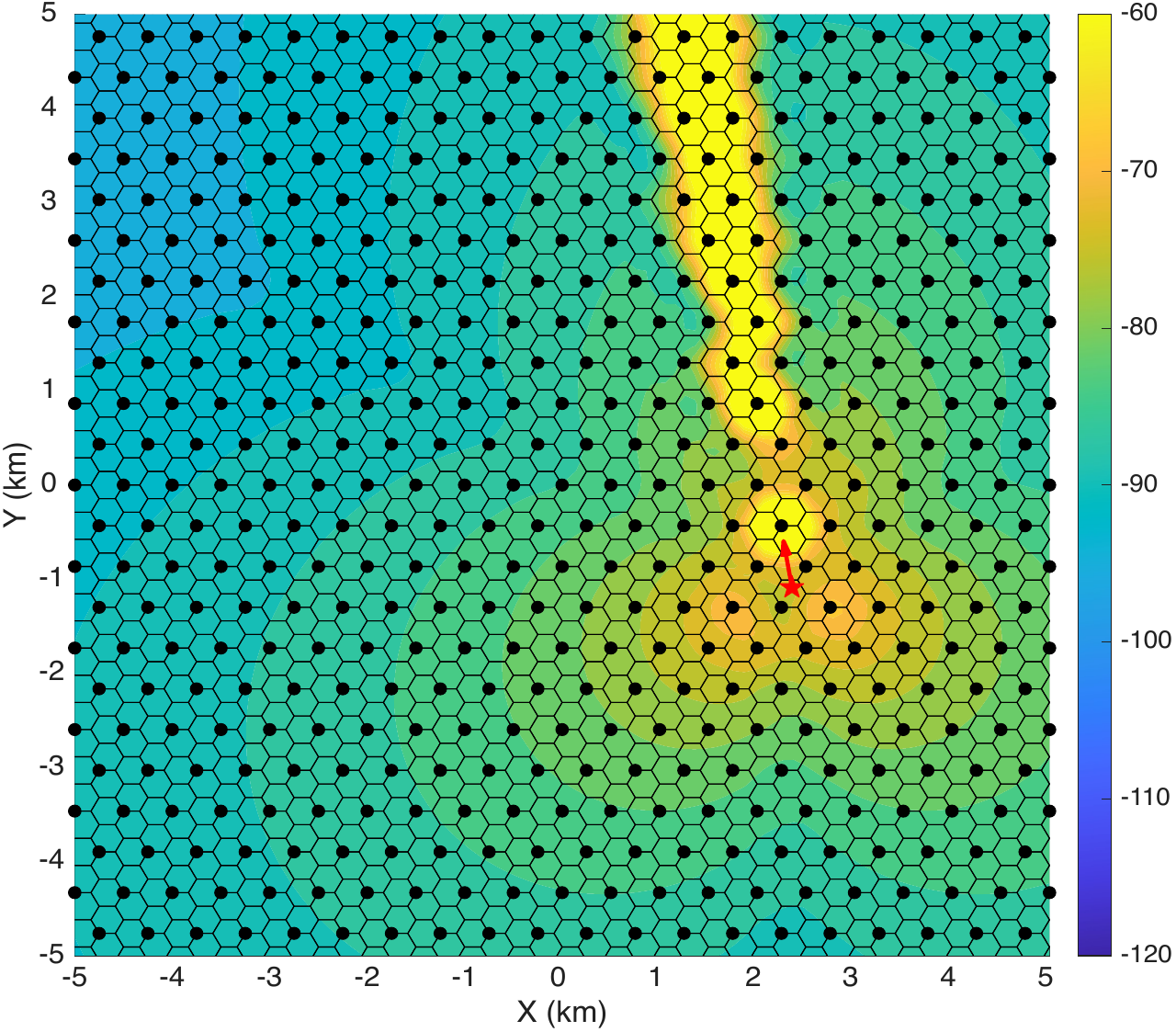}
  \end{minipage}
}\hfill
\subfloat[ISD = 1000 m]{%
  \begin{minipage}[t]{0.32\textwidth}
    \centering
    \includegraphics[width=\linewidth]{Figures/Figure2a_0102.pdf}
  \end{minipage}
}\hfill
\subfloat[ISD = 1500 m]{%
  \begin{minipage}[t]{0.32\textwidth}
    \centering
    \includegraphics[width=\linewidth]
    {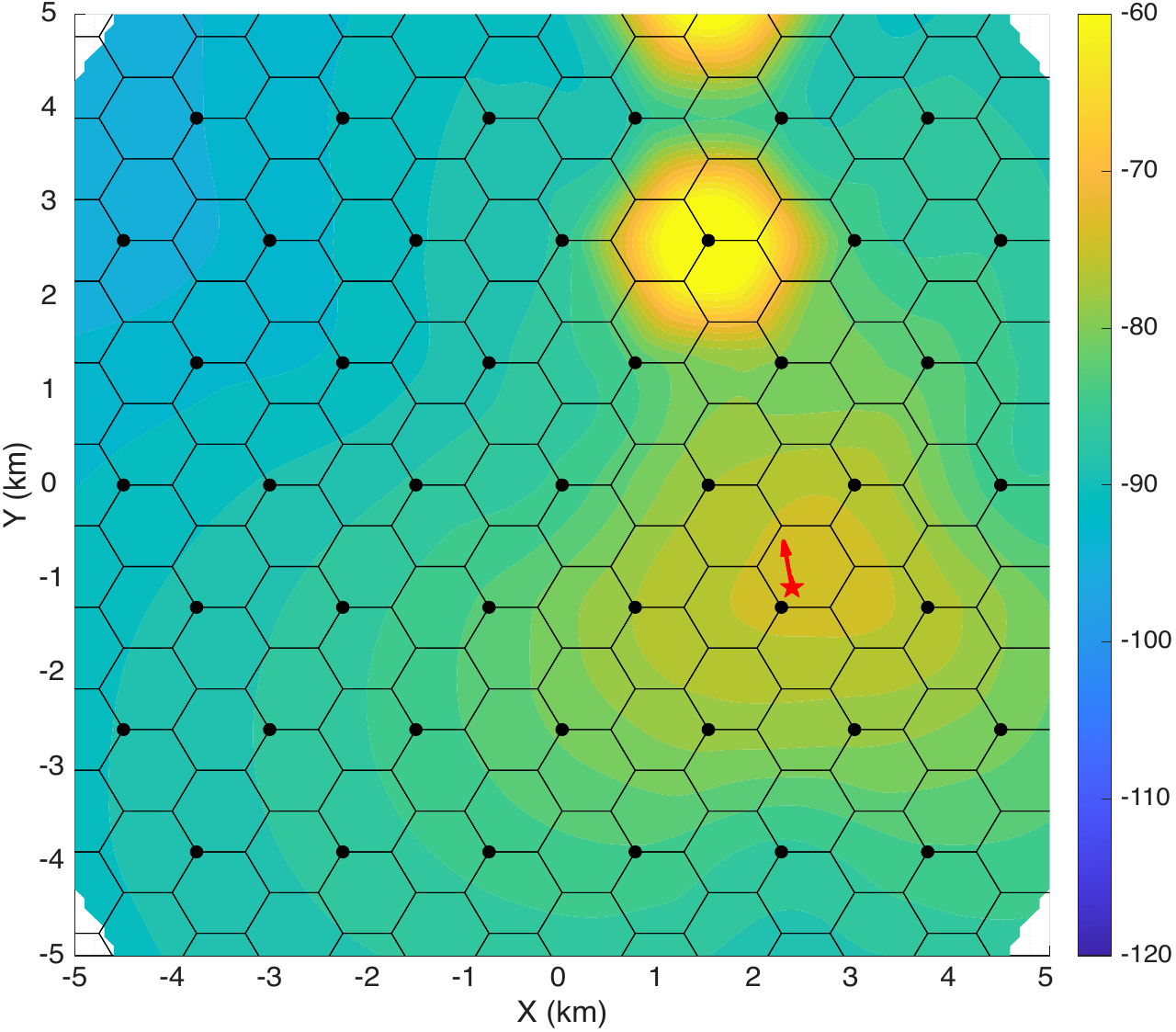}
  \end{minipage}
}

\caption{Spatial received-power maps under FSPL and different ISDs (From left to right: ISD = 500m, 1000m, and 1500m). The sensors use 3-sector antennas.}
\label{fig:heatmaps_ISD}
\end{figure*}

\section{Simulation Results}

This section evaluates the LarS-Net performance and the network-level sensing metrics introduced in Section~\ref{sec:metrics}. Unless otherwise stated, simulations are performed
over a $10~\mathrm{km}\times 10~\mathrm{km}$ area. The FS incumbent operates at carrier frequency $f_c = 7.25$~GHz with system bandwidth $B = 30$~MHz, $P_{\text{EIRP,max}}$ of $63$~dBm, and height $h_{\text{I}} = 60$~m. The FS incumbent transmit employs a commercial 3-foot diameter parabolic dish antenna,  modeled by a 3D directional pattern $G_{\text{I}}(\phi,\theta)$ with peak gain $G_{\text{I,max}} = 33.1\,\text{dBi}$. This high-gain, narrow-beam configuration corresponds to an approximate $3.7^\circ$ half-power beamwidth (HPBW) in both azimuth and elevation. A front-to-back ratio of $A_m = 40\,\text{dB}$ is applied, and the boresight azimuth is randomized by drawing an offset $\phi_{\text{I}}$. Moreover, measurement uncertainty is modeled by adding zero-mean Gaussian noise with a standard deviation of $3$~dB to the received power at each BS in every sensing slot.

Each BS is placed at a height $h_{\text{BS}}$ of 25 m and acts solely as a passive sensor. While most of modern BSs utilize arrays with at least 8$\times$4 elements, 3GPP is considering configurations for 6G urban macrocells with as much as 2304 elements near 7 GHz \cite{3gpp_tr_38_914}-\cite{3gpp_pcr_38_914}. For the integrated sensing framework in the simulation, we select an 8$\times$ vertical array. Following the 3GPP TR 38.921\cite{3gpp_tr_38_921} sectored antenna model with a per-element gain of 6.4 dBi, the array achieves a maximum gain of 15.4 dBi with a vertical HPBW of 9$^\circ$ and a horizontal HPBW of 90$^\circ$. For comparison, the co-located stand-alone sensor deployment utilizes an omnidirectional dipole antenna with a maximum gain of 7 dBi and a vertical HPBW of 18$^\circ$, the corresponding patter is calculated by the ITU-R F.1336 model~\cite{ITU1336}. To maintain consistency with established regulatory sensing standards, we adopt a sensing threshold of $-89$~dBm/MHz, mirroring the requirements for Environmental Sensing Capability (ESC) networks in the Citizen Broadband Radio Service (CBRS).

\subsection{Visualized Sensing Behavior vs. ISD}
We first analyze the spatial distribution of received incumbent power for three representative ISD values, namely $500$~m, $1000$~m, and $1500$~m. Fig.~\ref{fig:heatmaps_ISD} illustrates the received-power footprints under different
deployment densities and sectored antennas. As the ISD increases from $500$~m to
$1500$~m, the number of BSs illuminated by the incumbent decreases and the coverage becomes visibly sparser, reflecting weaker network sensing reach. Nevertheless, even at $1500$~m the network still captures a meaningful portion of the incumbent footprint, confirming that moderate densification is sufficient to maintain acceptable
sensing performance.

\subsection{Network-Level Sensing Metrics Across ISD}
\begin{figure}[t]
    \centering
    \includegraphics[width=1\columnwidth]{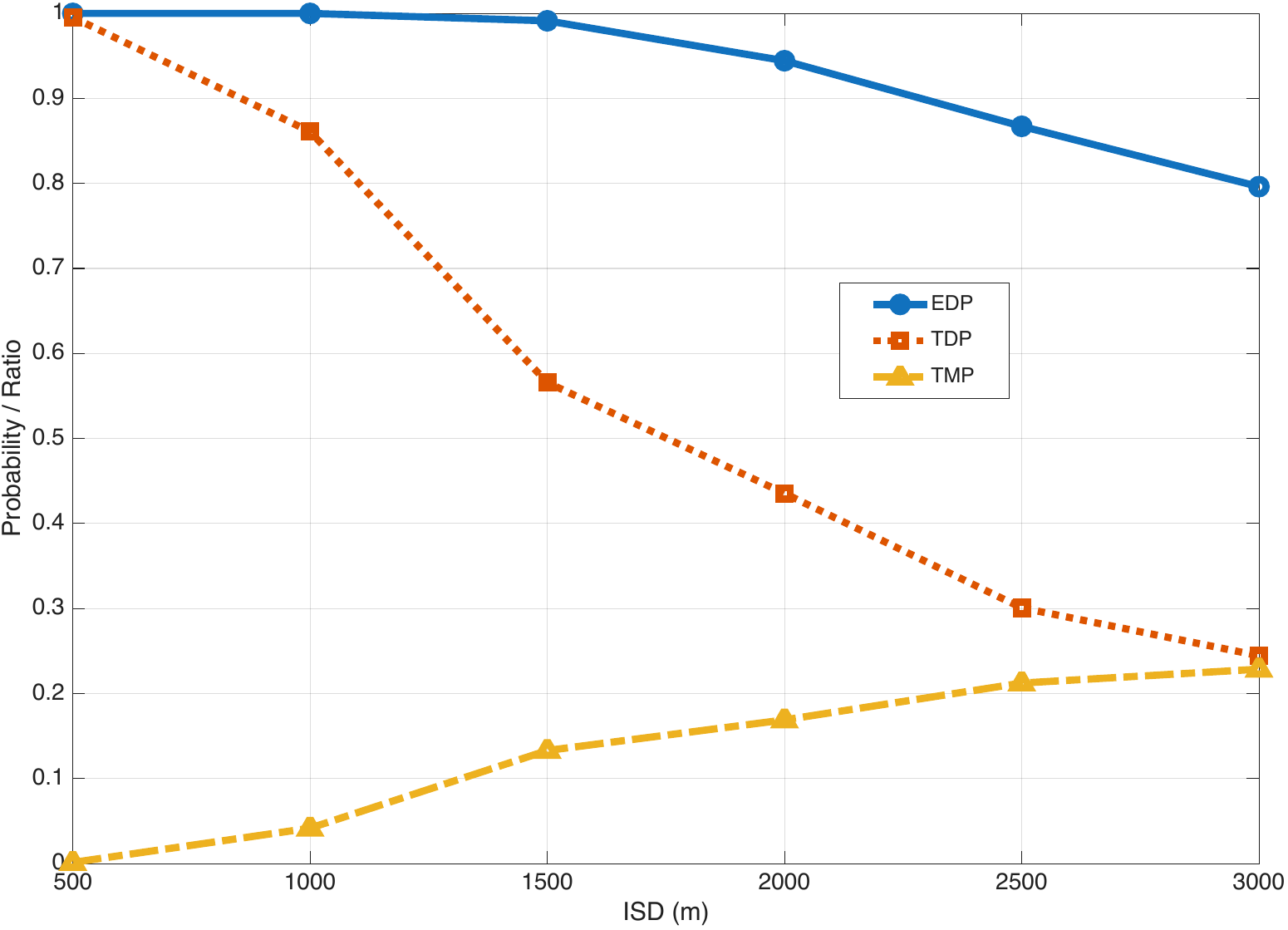}
    \caption{EDP, TDP, and TMP versus ISD under the FSPL and 3-sector antenna ($d_s = 0.2$, $p_{\text{on}} = 0.3$, noise = 3 dB).}
    \label{fig:EDP_lrm}
\end{figure}

Fig.~\ref{fig:EDP_lrm} presents the proposed network-level sensing metrics (EDP, TDP, and TMP) as functions of ISD under the FSPL. EDP is over 95\% when ISD is equal to or smaller than 2000 m, indicating that the network retains a very high probability of detecting incumbent emissions even under sparse deployments. Introducing the time-domain sensing model reveals the expected duty-cycle penalty: with $d_s=0.2$, TDP is consistently below EDP because only a fraction of BSs are active in each slot due to the consideration of the duty cycle, while TMP increases accordingly as opposed to TDP.

Table~\ref{tab:isd_90edp} summarizes the LarS-Net ISD required to achieve at least 90\% EDP across multiple antenna configurations and channel models (FSPL and LRM). Under the more realistic Longley-Rice model, the required LarS-Net ISD for an omnidirectional sensor is 1000 m. Compared to a standard cellular communication network ISD of 500 m, this implies that spectrum sensors only need to be deployed in every second BS in both $x$ (west-east) and $y$ (north-south) directions, corresponding to approximately 25\% of the total BS sites in the area. Furthermore, for the integrated sensing framework utilizing sectored antennas, the required LarS-Net ISD increases to 2500 m, meaning spectrum sensing capability is only required at 4\% of BS sites. Across all evaluated cases, the 90\%-EDP target is achieved with moderate densification (ISD on the order of $1$–$2.5$~km), rather than requiring very tight site spacing. Moreover, the performance gap between LRM and FSPL arises from the fact that FSPL represents a conservative free-space baseline, whereas LRM captures additional propagation mechanisms that are relevant for wide-area sensing at mid-band frequencies. These results collectively suggest that stringent incumbent detection can be achieved without overly dense sensing infrastructure, enabling substantial reductions in network deployment cost while remaining robust to both propagation-model and antenna-model choices.

\begin{table}[t]
\caption{LarS-Net ISD Required to Reach at Least 90\% EDP}
\label{tab:isd_90edp}
\centering
\begin{tabular}{|
>{\centering\arraybackslash}m{0.28\columnwidth}|
>{\centering\arraybackslash}m{0.25\columnwidth}|
>{\centering\arraybackslash}m{0.25\columnwidth}|
}
\hline
\multirow{2}{*}{Sensor Antenna}
& \multicolumn{2}{c|}{Channel Model} \\
\cline{2-3}
& FSPL &  LRM \\
\hline
Tri-sector & 2000 m &  2500 m \\
\hline
Omnidirectional & 1000 m & 1000 m \\
\hline
\end{tabular}
\end{table}

\begin{figure}[t]
    \centering
    \includegraphics[width=1\columnwidth]{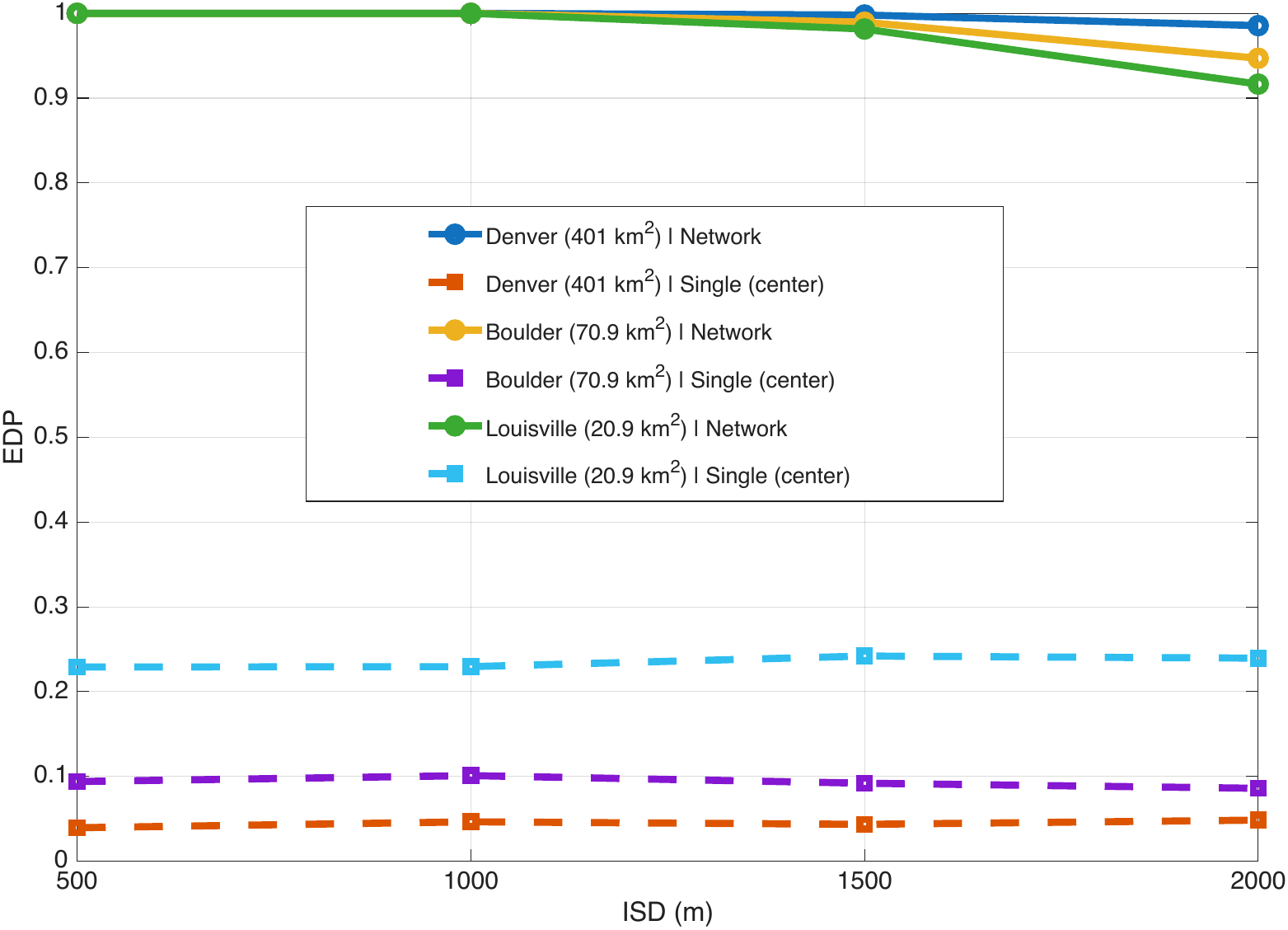}
    \caption{EDP versus ISD for three city footprints (areas indicated in the legend). ``Network'' denotes $K$-out-of-$N$ detection across all deployed BSs with $K=1$, while ``Single (center)'' denotes detection using the single sensor at the region center. Results are averaged over 2000 random incumbent drops per (ISD, city) under FSPL.}
    \label{fig:compare}
\end{figure}

\subsection{Performance Comparison of LarS-Net and Single Sensor}
As shown in Fig.~\ref{fig:compare}, we compare EDP of a multi-BS sensing network (LarS-Net) with a single-sensor baseline. For each ISD and each city footprint, we generate 2000 independent incumbent drops uniformly over the region (disk-shaped footprints with the areas indicated in the legend). For every drop, the received power is evaluated at all deployed BS sites under the same propagation and antenna assumptions (FSPL and 3-sector antenna). The network declares detection using a $K$-out-of-$N$ rule with $K=1$, i.e., an emission is detected if any BS site exceeds the per-MHz threshold. The single-sensor baseline uses only the BS at the region center.

Three areas of interest (AoIs) are compared in Fig.~\ref{fig:compare}: the large-scale urban environment Denver, Colorado, USA (401~km$^2$; population 735,000); a mid-size city of Boulder, Colorado, USA (71~km$^2$; population 110,000); and the smaller municipality of Louisville, Colorado, USA (21~km$^2$; population 21,000). As illustrated in the results, a single sensor cannot effectively detect an incumbent even in a small city and its detection performance degrades significantly as the size of AoI expands. In contrast, the LarS-Net framework remains a high EDP regardless of the AoI sizes, demonstrating the scalability and robustness of a distributed sensing architecture.

The network-level EDP remains high as ISD increases because detection relies on the presence of at least one BS with a sufficiently strong link to the incumbent. With a fixed ISD, enlarging the sensing region increases the total number of deployed BS sites, which in turn raises the probability that a randomly appearing incumbent is close to at least one sensor. As a result, the network benefits from increased spatial diversity and maintains strong detection performance even under sparser deployments.

In contrast, the single-sensor EDP remains low across all ISDs, since a center-located sensor has inherently limited spatial visibility. Incumbents that appear away from the center are unlikely to be detected, regardless of the overall network density. This disparity becomes more pronounced for larger regions (e.g., Denver), where the likelihood that an incumbent appears far from the center is higher. These results demonstrate that wide-area sensing performance is governed by network-level spatial diversity, and that single-sensor measurements substantially underestimate detectability over large or moderately sized regions.

\section{Discussion and Conclusions}

This paper introduced LarS-Net, a unified framework for analyzing large-scale distributed spectrum sensing and its integration with ISAC operations. By combining realistic cellular topology, directional incumbent emission, and physics-based propagation, the framework exposes the spatial mechanisms that determine sensing reliability at scale. The proposed network-level metrics: EDP, TDP, and TMP, provide a coherent means of quantifying detection performance and enable a consistent interpretation of how deployment density, duty-cycled operation, and noise uncertainty interact. By strategically selecting different subsets of cellular communication network sites, the analysis indicates that only enabling spectrum sensing capability on 4\% to 25\% of the existing infrastructure is sufficient to achieve over 90\% EDP. This result is economically promising for commercial deployment and long-term maintenance of future LarS-Net architecture. The comparison between network-wide sensing and a single-sensor receiver at the origin further illustrates that spatial diversity is indispensable. Taken together, LarS-Net provides both a practical simulation tool and an analytical foundation for sensing-oriented deployment planning and future 6G shared-spectrum system design. Future work will focus on extending the LarS-Net framework from homogeneous to heterogeneous topologies and evaluating detection performance for a broader range of incumbent systems. Additionally, we intend to refine the energy-detection thresholds for specific incumbent types and develop algorithms to dynamically optimize the sensing profile based on real-time environmental conditions.

\bibliography{reference}{}

@STRING{JSAC="{IEEE J. Sel. Areas Commun.}"}

@STRING{JSTSP="{IEEE J. Sel. Top. Signal Process.}"}

@STRING{TSP="{IEEE Trans. Signal Process.}"}

@STRING{IOT="{IEEE Internet Things J.}"}

@STRING{NW="{IEEE Netw.}"}

@ARTICLE{raouf2025,
  author={Raouf, Amir Hossein Fahim and Sun, Ruoyu and Poletti, Mark J.},
  journal=nw,
  title={{Spectrum usage monitoring and airtime utilization: Insights from a practical case study}}, 
  year={2025},
  volume={},
  number={},
  pages={1-1},
  doi={10.1109/MNET.2025.3571309}}

@ARTICLE{Bazerque2010,
  author={Bazerque, Juan Andrés and Giannakis, Georgios B.},
  journal=tsp, 
  title={{Distributed spectrum sensing for cognitive radio networks by exploiting sparsity}}, 
  year={2010},
  volume={58},
  number={3},
  pages={1847-1862},
  month     = mar,
  doi={10.1109/TSP.2009.2038417}}

@ARTICLE{Zeng2011,
  author={Zeng, Fanzi and Li, Chen and Tian, Zhi},
  journal=jstsp, 
  title={{Distributed compressive spectrum sensing in cooperative multihop cognitive networks}}, 
  year={2011},
month     = feb,
  volume={5},
  number={1},
  pages={37-48},
  doi={10.1109/JSTSP.2010.2055037}}

@article{Trigka2022,
  author    = {M. Trigka and E. Dritsas},
  title     = {{An efficient distributed approach for cooperative spectrum sensing in varying interests cognitive radio networks}},
volume={22},
  number={17},
  pages={6692},
  journal   = {Sensors},
  month     = sep,
  year      = {2022}
}

@inproceedings{Xu2023_ISAC_DAN,
  author    = {D. Xu and A. Khalili and X. Yu and D. W. Kwan Ng and R. Schober},
  title     = {{Integrated sensing and communication in distributed antenna networks}},
  booktitle = {Proc. IEEE ICC Workshops},
  address   = {Rome, Italy},
  year      = {2023}
}

@techreport{3gpp_tr_22_870,
  author      = "{3GPP}",
  title       = "{Study on 6G Use Cases and Service Requirements}",
  institution = "3rd Generation Partnership Project (3GPP)",
  type        = "TR",
  number      = "22.870",
  version     = "17.2.0",
  year        = "2025",
  month       = dec
}

@techreport{3gpp_tr_38_914,
  author      = "{3GPP}",
  title       = "{Study on 6G Scenarios and requirements}",
  institution = "3rd Generation Partnership Project (3GPP)",
  type        = "TR",
  number      = "38.914",
  version     = "0.2.0",
  year        = "2025",
  month       = dec
}

@techreport{3gpp_pcr_38_914,
  author      = "{3GPP}",
  title       = "{pCR to TR 38.914 v0.2.0 with RAN \#110 results for deployment scenarios}",
  institution = "3rd Generation Partnership Project (3GPP)",
  type        = "",
  number      = "RP-253873",
  version     = "",
  year        = "2025",
  month       = dec
}

@techreport{3gpp_tr_38_921,
  author      = "{3GPP}",
  title       = "{Study on International Mobile Telecommunications (IMT) parameters for 6.425 - 7.025 GHz, 7.025 - 7.125 GHz and 10.0 -  10.5 GHz}",
  institution = "3rd Generation Partnership Project (3GPP)",
  type        = "TR",
  number      = "38.921",
  version     = "19.0.0",
  year        = "2025",
  month       = sep
}

@techreport{ITU1336,
  author      = "{International Telecommunication Union – Radiocommunication Sector (ITU-R)}",
  title       = "{Reference peak and average patterns for antennas of digital fixed wireless systems to be used in sharing studies in the frequency range from 1 GHz to about 70 GHz}",
  type        = "Recommendation ITU-R F.1336-5",
    institution = "ITU",
  number      = "",
  version     = "",
  year        = "2019",
}

@INPROCEEDINGS{sun2024icc,
  author={Sun, Ruoyu and Yi, Yunjung and Poletti, Mark},
  booktitle={Proc. IEEE Int. Conf. Commun. Workshops}, 
  title={{Base station antenna array size impact on interference between TDD cellular networks}}, 
  year={2024},
  volume={},
  number={},
address = {Denver, CO, USA},
  pages={2107-2112},
  doi={10.1109/ICCWorkshops59551.2024.10615366}}

@ARTICLE{ramsha2022_HybridML_IoV,
  author={Ahmed, Ramsha and Chen, Yueyun and Hassan, Bilal and Du, Liping and Hassan, Taimur and Dias, Jorge},
  journal=iot,
  title={{Hybrid machine-learning-based spectrum sensing and allocation with adaptive congestion-aware modeling in CR-assisted IoV networks}}, 
  year={2022},
month =dec,
  volume={9},
  number={24},
  pages={25100-25116},
  doi={10.1109/JIOT.2022.3195425}}

@ARTICLE{zhang2024cross,
  author={Zhang, Min and Zhu, Xiaoying and Zhang, Bo and Wang, Shi and Sun, Hao},
  journal={IEEE Sens. J.}, 
  title={{A cross-layer performance evaluation system for spectrum sensing and allocation strategies in CR-WSN}}, 
  year={2024},
month = may,
  volume={24},
  number={9},
  pages={15355-15366},
  doi={10.1109/JSEN.2024.3367758}}

@INPROCEEDINGS{zang2023spectrum,
  author={Zang, Junwei and Liu, Qiao and He, Jia and Wang, Guangjian},
  booktitle={Proc. IEEE Veh. Technol. Conf. (VTC2023-Spring)}, 
  title={{On spectrum sensing for mmWave and THz beam-based communications}}, 
  year={2023},
  volume={},
  number={},
  pages={1-6},
address = {Florence, Italy},
  doi={10.1109/VTC2023-Spring57618.2023.10199957}}

@ARTICLE{Liu2022_ISAC_Mag,
  author={Liu, Fan and Cui, Yuanhao and Masouros, Christos and Xu, Jie and Han, Tony Xiao and Eldar, Yonina C. and Buzzi, Stefano},
  journal=jsac, 
  title={{Integrated sensing and communications: Toward dual-functional wireless networks for 6G and beyond}}, 
  year={2022},
month =jun,
  volume={40},
  number={6},
  pages={1728-1767},
  doi={10.1109/JSAC.2022.3156632}}

@ARTICLE{guo2025dmimo,
  author={Guo, Hao and Wymeersch, Henk and Makki, Behrooz and Chen, Hui and Wu, Yibo and Durisi, Giuseppe and Keskin, Musa Furkan and Moghaddam, Mohammad H. and Madapatha, Charitha and Yu, Han and Hammarberg, Peter and Kim, Hyowon and Svensson, Tommy},
  journal={IEEE Wireless Commun.}, 
  title={{Integrated communication, localization, and sensing in 6G D-MIMO networks}}, 
  year={2025},
month = apr,
  volume={32},
  number={2},
  pages={214-221},
  doi={10.1109/MWC.007.2400117}}

\bibliographystyle{IEEEtran}

\end{document}